\documentclass[aps, prd, amsmath, floats, floatfix, twocolumn,
superscriptaddress, nofootinbib, showpacs]{revtex4}
\usepackage{graphicx}
\usepackage{epsfig}
\usepackage{color}
\usepackage{soul}
\usepackage{url}
\usepackage{bm}         
\usepackage{times}

\newcommand{\beq}{\begin{equation}}
\newcommand{\eeq}{\end{equation}}
\newcommand{\beqn}{\begin{eqnarray}}
\newcommand{\eeqn}{\end{eqnarray}}

\usepackage{color}

\newcommand{\CITA}{\affiliation{Canadian Institute for Theoretical 
    Astrophysics, University of Toronto, Toronto, Ontario M5S 3H8, Canada}}

\usepackage{graphicx}
\usepackage{dcolumn}
\usepackage{bm}
\usepackage{epsf}

\begin{document}

\title{Black Hole-Neutron Star Mergers: Disk Mass Predictions}

\author{Francois Foucart} \CITA%

\begin{abstract}

Determining the final result of black hole-neutron star mergers, and in
particular the amount of matter remaining outside the black hole at late times and
its properties, has been
one of the main motivations behind the numerical simulation of these systems. 
Black hole-neutron star binaries are amongst the most likely progenitors of short 
gamma-ray bursts --- as long as massive (probably a few percents of a solar mass), hot
accretion disks are formed around the black hole. 
Whether this actually happens strongly depends on the physical
characteristics of the system, and in particular on the mass ratio, the spin of the black
hole, and the radius of the neutron star. We present here a simple two-parameter model,
fitted to existing numerical results, for the determination of the mass remaining outside 
the black hole a few milliseconds after a black hole-neutron star merger (i.e. the combined mass
of the accretion disk, the tidal tail, and the potential ejecta).
This model predicts the remnant mass within a few percents of the mass of the neutron star, 
at least for remnant masses up to $20\%$ of the neutron star mass. Results across the range of parameters
deemed to be the most likely astrophysically are presented here. We find that, for $10M_\odot$ black 
holes, massive disks are only possible for large neutron stars ($R_{\rm NS}\gtrsim 12{\rm km}$), or quasi-extremal 
black hole spins ($a_{\rm BH}/M_{\rm BH} \gtrsim 0.9$).
We also use our model to discuss how the equation of state of the neutron star affects the final 
remnant, and the strong influence that this can have on the rate of short gamma-ray bursts 
produced by black hole-neutron star mergers.

\end{abstract}


\maketitle

\section{Introduction}
\label{intro}

The potential of black hole-neutron star (BHNS) mergers as progenitors of short gamma-ray 
bursts (SGRBs) and their importance as sources of gravitational waves detectable by ground-based
interferometers such as Advanced LIGO, VIRGO, and 
KAGRA~\cite{LIGO,VIRGO,2010CQGra..27q3001A,2012CQGra..29l4007S},
have driven most recent studies of these systems.
Gamma-ray bursts, in particular, are a likely result if the neutron star is tidally disrupted, 
and the final outcome of the merger is a massive accretion disk around the black hole 
(see~\cite{2007NJPh....9...17L} and references therein). If the disruption of the neutron
star causes unbound material to be ejected from the system, radioactive decay in the
neutron-rich ejecta could also produce a 'kilonova', visible as a day-long, mostly isotropic
optical transient~\cite{2011ApJ...736L..21R,2012ApJ...746...48M}.

Numerical simulations have taught us that BHNS mergers can be divided 
into two broad categories: either tidal effects are strong enough for the neutron
star to be disrupted before reaching the innermost stable circular orbit (ISCO) of the
black hole, or the neutron star plunges into the hole before tidal disruption occurs. 
In the first case, some material from the disrupted star
remains outside the black hole for long periods of time ($\sim 0.1-1s$) in the form
of an accretion disk, a tidal tail, and/or unbound ejecta. In the second case, however,
the entire neutron star is rapidly accreted onto the black hole.
To first order, the most important parameters
determining the outcome of a BHNS merger are the mass ratio of the 
binary~\cite{Etienne:2008re,PhysRevD.84.064018,2012PhRvD..85d4015F}, the spin
magnitude of the black hole~\cite{Etienne:2008re,2011PhRvD..83b4005F,2012PhRvD..85d4015F}, 
its orientation~\cite{2011PhRvD..83b4005F}, and the size of the neutron 
star~\cite{Duez:2010a,2010PhRvD..82d4049K,PhysRevD.84.064018}. The formation
of massive accretion disks is more likely to occur for black holes of low mass
(at least down to mass ratios $M_{\rm BH}/M_{\rm NS}\sim 3$) and high spins, and for large neutron stars. 

Studying these mergers is a complex problem, and accurate results can
only be obtained through numerical simulations in a general relativistic framework: 
results using approximate treatments of gravity can lead to qualitative differences
in the dynamics of the merger, and large errors in the mass of the
final accretion disk or of any unbound material. Unfortunately, general relativistic 
simulations are computationally expensive, and only $\sim 50$ BHNS mergers have been
studied so far (see~\cite{2010CQGra..27k4002D,lrr-2011-6} for reviews of these results). 
Additionally, a majority of these simulations considered binaries with mass ratios
$M_{\rm BH}/M_{\rm NS}\sim 2-3$, while population synthesis models indicate that mass ratios 
$M_{\rm BH}/M_{\rm NS}\geq 5$ are astrophysically more 
likely~\cite{2008ApJ...682..474B,2010ApJ...715L.138B}. Existing general relativistic 
simulations are also fairly limited in the physical effects considered: only a few include magnetic 
fields~\cite{2010PhRvL.105k1101C,2012PhRvD..85f4029E,2012arXiv1209.1632E} or nuclear theory-based equations of 
state~\cite{Duez:2010a}, and none have considered neutrino emission (although neutrinos have
been included in simulations of neutron star-neutron star mergers~\cite{2012CQGra..29l4003K}).
Magnetic fields and neutrino radiation are unlikely to affect the disruption of the star, or the
amount of matter remaining outside the black hole after merger.
Magnetic fields exceeding $10^{17}G$ are necessary for the pre-merger evolution of the binary to be
modified~\cite{2012PhRvD..85f4029E}, while the neutron star had more than enough time to cool down
during the long evolution of the binary towards merger, so that neutrino emission over the short timescale
governing the disruption of the star ($\tau_{\rm dis}\sim 1{\rm ms}$) or even the last few orbits of evolution
($\tau_{\rm orbit}\sim 10{\rm ms}$) is negligible (see~\cite{2012CQGra..29l4003K} for a numerical
confirmation in the case of binary neutron star mergers).  
On the other hand, both effects are critical to the evolution of the post-merger remnant, and to the modelling of 
electromagnetic and neutrino counterparts to the gravitational wave signal emitted by 
black hole-neutron star mergers: accretion disks resulting from these mergers are
expected to be susceptible to the magneto-rotational instability, and cooled by
neutrino emission over a timescale $\tau_{\nu}\sim 0.1{\rm s}$~\cite{Lee:2005se} comparable
with the lifetime of the disk.

Given the size of the parameter space to explore and the cost of numerical simulations, obtaining
accurate predictions for the final state of the system for all possible configurations
is only feasible through the construction of a model which effectively interpolates 
between known numerical results. Such a model can also be of great help to determine which binary
parameters should be used in numerical simulations in order to study a specific physical effect 
(e.g. massive disks) without having to run many different configurations.
In the limit of extreme mass ratios ($M_{\rm NS} << M_{\rm BH}$), analytical expressions
can be obtained for the binary separation at which a neutron star would be disrupted
by the tidal field of a Kerr black hole~\cite{1973ApJ...185...43F,2000ApJ...532..530W}, and compared
with the innermost stable circular orbit (ISCO) of the hole to obtain a criteria separating binaries
which disrupt outside the ISCO from binaries for which the neutron star will directly plunge into the
black hole. A similar criteria for more symmetric mass ratios was derived analytically by
Miller~\cite{2005ApJ...626L..41M} using a Post-Newtonian approximation to the location of the ISCO due to
Damour et al.~\cite{2000PhRvD..62h4011D}, and numerically by 
Taniguchi et al.~\cite{2008PhRvD..77d4003T} for the case of non-spinning black holes
by studying the quasi-equilibrium configurations used as starting point for the numerical
evolution of BHNS binaries in general relativity. We discuss these approximations in 
Sec.~\ref{sec:prevmod}, and how they compare to our fit to recent numerical simulations.
More recently, Pannarale et al.~\cite{2011ApJ...727...95P} computed estimates for the mass 
remaining outside the black hole at late times through the use of a toy-model studying 
the tidal forces acting on the neutron star, represented by a tri-axial ellipsoid, and fitted
to the results of numerical simulations. However, many of those simulations underestimated 
the remnant masses, and only covered the low mass ratio regime
$M_{\rm BH}/M_{\rm NS}\sim 2-3$. The qualitative dependence of the remnant mass in the 
parameters of the binary is captured by their model, but the quantitative results do not 
match more recent simulations, particularly for larger black hole masses~\cite{2012PhRvD..85d4015F}. 

In this paper we show that simple models comparing the estimated 
separation at which tidal disruption of the neutron star occurs ($d_{\rm tidal}$) 
and the radius of the ISCO ($R_{\rm ISCO}$) can accurately predict the mass remaining 
outside the black hole at late times. We fit
four such models (with different approximations for $d_{\rm tidal}$) 
to a set of 26 recent numerical simulations covering mass ratios in the
range $M_{\rm BH}/M_{\rm NS} = 3-7$, black hole spins up to $0.9$ and neutron star radii 
$R_{\rm NS}\approx 11-16$ km.
The case of black hole spins misaligned with the orbital angular momentum is not
considered here, and we limit ourselves to low eccentricity orbits (high eccentricities only
occur when the binary is formed through dynamical capture, e.g. in nuclear or globular clusters). 
All models match the simulation
results within their expected numerical errors, a few percents of the original mass
of the neutron star. 

As obtaining simple approximate constraints on the binary parameters for which short gamma-ray bursts
might be produced is one of the potential use of this model, we will begin by summarizing in 
Sec.~\ref{sec:sgrb} the 
main channels through which BHNS mergers could generate such bursts, and discuss in this context
what can be learnt from a simple model predicting solely the total amount
of mass remaining outside the black hole a few milliseconds after merger.
We then describe in Sec.~\ref{sec:model} the models used, and 
their physical inspiration. Sec.~\ref{sec:NR} summarizes the numerical results
used to calibrate the models, while Sec.~\ref{sec:params} gives the best-fit
parameters, and discuss the quality of the fits. Finally, in Sec.~\ref{sec:discussion},
we show predictions of the simplest model across the entire parameter space. We also
discuss their strong dependence in the size of the neutron star, and potential
implications for the rate of short gamma-ray bursts originating from BHNS mergers. 

\section{Short Gamma-Ray Bursts}
\label{sec:sgrb}

One of the most interesting aspect of black hole-neutron star mergers is their potential
as progenitors of short gamma-ray bursts (SGRBs) --- a potential which is however conditional on
their ability to form massive hot disks around the remnant black hole.
A detailed discussion of the characteristics of SGRBs is beyond the scope
of this article. But in order to better understand the implications of our model for the production
of SGRBs, a few relevant characteristics and potential pathways to SGRBs should be summarized.
The interested reader can find more details in, for example, the review of SGRBs
progenitors by Lee \& Ramirez-Ruiz~\cite{2007NJPh....9...17L}. SGRBs are extremely energetic
events, releasing energies $E\sim 10^{48-51} (\Omega/4\pi){\rm ergs}$ over a duration varying between a few 
milliseconds and a few seconds (where $\Omega$ is the solid angle over which the energy is
emitted). As opposed to long bursts, which are observed in star forming regions of galaxies and
whose association with core-collapse supernovae is generally accepted, the origin of SGRBs 
remains controversial. SGRBs are observed in all types of galaxies, including in regions without
significant star formation. And some of them even appear offset with respect to their most likely host. 
Compact mergers are thus a tantalizing option as SGRBs progenitors: they could release
the required energies, they occur long after star formation, and a velocity kick given to a neutron
star during an asymmetric supernova explosion could explain an offset with respect to the host
galaxy.

Two main pathways have been proposed to get to a SGRB from the remnant of a BHNS (or binary
neutron star) merger. The first involve the emission by a hot accretion disk of neutrinos and anti-neutrinos,
which can recombine in high-energy electron-positron pairs in a baryon free region along the spin axis of the
central black hole, driving an ultra-relativistic wind~\cite{1993Natur.361..236M}. 
Determining the energy emitted is a complex problem,
depending on the efficiency of the conversion of the fluid energy into neutrino radiation, the efficiency
of the $\nu\tilde\nu \rightarrow e^- e^+$ recombination and the creation of a region sufficiently free
of matter to allow the production of an ultra-relativistic, collimated outflow. Two dimensional disk
simulations indicate that, for a disk density $\rho\sim 10^{10-11}{\rm g/cm^3}$ and temperature $T\sim 2-5{\rm MeV}$, an energy
output $E\approx 10^{49}(m_d/0.03M_\odot)^2$ can be expected, with $m_d$ the mass of the
accretion disk~\cite{2005ApJ...630L.165L}. Another possibility is to extract the rotational energy of
the black hole through electromagnetic torques (Blandford-Znajek mechanism~\cite{1977MNRAS.179..433B}). 
This requires the rapid growth of a large
poloidal magnetic field, to roughly equipartition levels. Whether this occur in practice remains an
open question. Assuming equipartition of energy, Lee et al.~\cite{2005ApJ...630L.165L} find
that an energy $E\approx 5\times 10^{50}(m_d/0.03M_\odot)(\alpha/0.1)^{-0.55}$ is released
(and $E$ scales like $B_p^2$ for magnetic energies below equipartition). Here, $\alpha$ is
the viscosity of the disk, and $B_p$ the poloidal field. 

From this brief summary, we can see that the physics governing the generation of SGRBs is complex,
and not entirely understood. Accordingly, it would be impossible to determine whether
a SGRB can be produced from a BHNS merger simply from the total mass remaining outside of the black
hole at late times. This mass is, however, an important indicator of what happened during merger,
and of the energy available for post-merger evolution. Typically, numerical simulations show that when 
the remnant mass is greater than $\sim 0.1M_\odot$ about $1/3-2/3$ of that
mass is in a disk, and the rest in the tidal tail. The temperature and density of the disk are generally
compatible with the assumptions of Lee et al.~\cite{2005ApJ...630L.165L}, except for the lower
mass disks around black holes $\geq 10M_\odot$, which have fairly low densities. This
seems like a promising setup. But without a better understanding of the exact conditions leading to
the production of a SGRB, we cannot know for sure which of these configurations, if any, would be SGRB
progenitors. For lower remnant mass, the situation is more parameter dependent: for lower black
hole masses, the formation of a hot accretion disk remains possible, while for higher mass ratios, 
or when the black hole spin is strongly misaligned with the orbital angular momentum, nearly all of the material
is sent in an elongated tidal tail.
In the end, the only certainty comes for configurations in which no matter remains outside of the black
hole: these cases can certainly be excluded as SGRB progenitors --- and this already
rules out a significant part of the BHNS parameter space.

\section{Tidal Disruption Models}
\label{sec:model}

The models used here to estimate the mass remaining outside the black hole at late times
are based on a comparison between the binary separation at which tidal forces become
strong enough to disrupt the star, $d_{\rm tidal}$, and the radius of the innermost stable
circular orbit $R_{\rm ISCO}$. Intuitively, if $d_{\rm tidal}\lesssim R_{\rm ISCO}$, the neutron
star will plunge directly into the black hole and no mass will remain outside the hole
after merger. On the other hand, if $d_{\rm tidal} \gtrsim R_{\rm ISCO}$, the star will be disrupted.
Some disrupted material will then form an accretion disk, while some will be ejected in a tidal tail
and fall back on the disk over timescales long with respect to the duration of the merger
(most of the neutron star material is accreted within a few milliseconds, and the disk settles
to a near equilibrium profile over $\sim 10{\rm ms}$, while material in the tidal tail will fall
back over longer timescales $\sim 0.1-1{\rm s}$). Finally, it is possible that up to a few
percents of the neutron star material will be unbound.

The separation $d_{\rm tidal}$ at which tidal disruption occurs can be estimated in Newtonian 
theory by balancing the gravitational acceleration due to the star with the tidal acceleration due 
to the black hole:
\beqn
&& \frac{M_{\rm NS}}{R_{\rm NS}^2} \sim \frac{3M_{\rm BH}}{d_{\rm tidal}^3} R_{\rm NS}\\
&&\label{eq:dtid} d_{\rm tidal} \sim R_{\rm NS} \left(\frac{3M_{\rm BH}}{M_{\rm NS}}\right)^{1/3},
\eeqn
where $R_{\rm NS}$ is the radius of the neutron star, $M_{\rm NS}$ and $M_{\rm BH}$ are
the masses of the compact objects, and we work in units in which $G=c=1$. 
In general relativity, these quantities are not uniquely defined.
In practice we will use the radius of the star in Schwarzschild coordinates and the ADM mass
of the compact objects, all measured at infinite separation. 
\footnote{
The numerical factor
of '$3$' is chosen to match more closely the results of Fishbone~\cite{1973ApJ...185...43F} used
in the 'Kerr' model~(\ref{eq:modtil})  --- but is practically of no importance here,
as  we only rely on the scaling of $d_{\rm tidal}$ with the binary parameters in our model.
}

As for the radius of the ISCO, it is given by \cite{1972ApJ...178..347B} 
\beqn
Z_1 & = & 1+(1-\chi_{\rm BH}^2)^{1/3}\left[(1+\chi_{\rm BH})^{1/3} + (1-\chi_{\rm BH})^{1/3}\right] \nonumber\\
Z_2 & = & \sqrt{3\chi_{\rm BH}^2+Z_1^2} \nonumber\\
\frac{R_{\rm ISCO}}{M_{\rm BH}} &=& 3+Z_2-{\rm sign}(\chi_{\rm BH})\sqrt{(3-Z_1)(3+Z_1+2Z_2)}
\eeqn
where $\chi_{\rm BH}=a_{\rm BH}/M_{\rm BH}$ is the dimensionless spin parameter of the black hole.

To construct a model for the fraction of the baryon mass of the star remaining outside the black hole 
at late times, we assume that this mass is entirely determined by the relative location
of $R_{\rm ISCO}$ and $d_{\rm tidal}$, in units of the neutron star radius. A first guess for the remnant mass
$M^0_{\rm model}$ is then the linear model:
\beq
\label{eq:M0}
\frac{M^0_{\rm model}}{M^b_{\rm NS}}=\alpha^0 \frac{d_{\rm tidal}}{R_{\rm NS}}-\beta^0 \frac{R_{\rm ISCO}}{R_{\rm NS}}+\gamma_0,
\eeq
where $\alpha^0$, $\beta^0$ and $\gamma_0$ are the free parameters of the model, and $M^b_{\rm NS}$ is the 
baryon mass of the neutron star. However, this simple prescription fits the numerical
data rather poorly. In particular, it stronly underestimates the impact of the neutron star compactness 
$C_{\rm NS}=M_{\rm NS}/R_{\rm NS}$ on the result. This problem is not overly surprising: $d_{\rm tidal}$ was derived in
Newtonian gravity, but applied to compact objects. In particular, it predicts a finite radius for tidal
disruption even if we replace the neutron star by a non-spinning black hole (for which $C=0.5$). 
To improve the model, we use instead a corrected estimate of the distance for tidal disruption,
in which compact objects are more strongly bound:
\beq
\tilde d_{\rm tidal} = d_{\rm tidal}(1-2C_{\rm NS}).
\eeq
This leads to the following model for the mass remaining outside the black hole at late times, $M^{\rm rem}_{\rm model}$:
\beq
\label{eq:mod}
\frac{M^{\rm rem}_{\rm model}}{M^b_{\rm NS}}= \alpha \left(3q\right)^{1/3}(1-2C_{\rm NS}) 
- \beta \frac{R_{\rm ISCO}}{R_{\rm NS}},
\eeq
with $q=\frac{M_{\rm BH}}{M_{\rm NS}}$.
We could have added a constant term $\gamma$ as in Eq.~\ref{eq:M0}, but find that this does not
improve the quality of the fit. At the current level of accuracy of numerical simulations, we will show in 
Sec.~\ref{sec:params} that this simple model is in agreement with known results.
We should note that, as written here, Eq.~\ref{eq:mod} can predict negative remnant mass. These should be
understood as the absence of any matter outside of the black hole after merger, i.e. $M^{\rm rem}=0$.

A potential improvement on the model described by Eq.~\ref{eq:mod} is to compute the tidal effects from the Kerr metric
instead of the Newtonian formula. Fishbone~\cite{1973ApJ...185...43F} obtained analytical results for these effects.
Using his results leads to a correction to the value of the separation at which tidal disruption occurs: 
$\xi_{\rm tidal}=d_{\rm tidal}/R_{\rm NS}$ is then solution of the implicit equation
\beq
\frac{M_{\rm NS}\xi_{\rm tidal}^3}{M_{\rm BH}}=\frac{3(\xi_{\rm tidal}^2-2\kappa \xi_{\rm tidal}+ \chi_{\rm BH}^2\kappa^2)}
{\xi_{\rm tidal}^2-3\kappa \xi_{\rm tidal}+2\chi_{\rm BH}\sqrt{\kappa^3\xi_{\rm tidal}} }
\eeq 
with $\kappa=M_{\rm BH}/R_{\rm NS}$. We can then write the corrected model
\beq
\label{eq:modtil}\frac{\tilde M^{\rm rem}_{\rm model}}{M^b_{\rm NS}} = 
\tilde \alpha \xi_{\rm tidal}(1-2C_{\rm NS})-\tilde \beta \frac{R_{\rm ISCO}}{R_{\rm NS}}.
\eeq
In practice, $\tilde M^{\rm rem}_{\rm model}$ gives results consistent with the simpler model $M^{\rm rem}_{\rm model}$.

In both cases, we end up with a simple formula for the predicted fraction of the neutron star mass remaining outside of
the black hole at late times as a function of only 3 dimensionless parameters:
the mass ratio $q$, the neutron star compactness $C_{\rm NS}$ and the dimensionless spin of the BH $\chi_{\rm BH}$.
Clearly, these are not enough to entirely determine the characteristics of the binary: the total mass of the system as
well as the internal structure of the neutron star are required to do so. The structure of the star, in particular, is
expected to affect the remnant mass - although not as much as its compactness. At best, these models can thus only
be accurate up to variations in the remnant mass due to changes in the properties of the neutron star matter that do 
not modify $C_{\rm NS}$ (see Sec.~\ref{sec:error} for a more detailed discussion of the accuracy of the model).

Determining which characteristics of the star are probed by a study of its disruption is in fact a complex problem.
In the Newtonian, extreme mass ratio limit, and for polytropic equations of state ($P=\kappa \rho^{1+1/n}$), the tidal
disruption radius is proportional to $k_2^{1/3} (1-n/5)^{1/3} R_{\rm NS} q^{1/3}$, where $k_2$ is the tidal
Love number of the neutron star~\cite{1993ApJS...88..205L}. In general relativity and for more
symmetric mass ratios, this expression will however be modified. Additionally, the location of the ISCO itself
depends on the properties of the star. For non-spinning black holes and $n=1$ polytropes, this dependence was estimated by Taniguchi et al.~\cite{2008PhRvD..77d4003T}. All these
physical effects are not taken into account in our model. In a way, they are what we fit for when
we choose the free parameters $\alpha$ and $\beta$. This complex picture can be contrasted with the more
simple interpretation of the effect of tides on the gravitational waveforms during a BHNS inspiral, which
causes an accumulated phase difference in the signal proportional to 
$k_2 R^5_{\rm NS}$~\cite{2008PhRvD..77b1502F,2010PhRvD..81l3016H} (at the lowest order at which finite-size effects
enter post-newtonian approximations to the gravitational wave signal). 

Models which are theoretically as valid
as~(\ref{eq:mod}) and fit the data as well can easily be built by including some of those corrections. For example,
including the Newtonian dependence of $d_{\rm tidal}$ in the dimensionless Love number $k_2$ gives
\beq
\label{eq:mod2}
\frac{M^{\rm rem}_{\rm model,k}}{M^b_{\rm NS}}=  0.534\left(3k_2q\right)^{1/3}(1-2\frac{M_{\rm NS}}{R_{\rm NS}}) 
- 0.119 \frac{R_{\rm ISCO}}{R_{\rm NS}},
\eeq
while a model using as input parameters the quantity that can most easily be measured in gravitational
wave signals $\rho_{\rm NS}=(k_2/0.1)^{1/5} R_{\rm NS}$ (which could be directly compared with the
results of a gravitational wave measurement of the neutron star properties) can be written as
\beq
\label{eq:mod3}
\frac{M^{\rm rem}_{\rm model,\rho}}{M^b_{\rm NS}}= 0.262 \left(3q\right)^{1/3}(1-2\frac{M_{\rm NS}}{\rho_{\rm NS}}) 
- 0.128 \frac{R_{\rm ISCO}}{\rho_{\rm NS}}.
\eeq
The normalization of $0.1$ for $k_2$ is arbitrary, and chosen to lie in the middle of the range of values
covered by simulations ($k_2=0.085-0.135$, as given in Hinderer~\cite{2008ApJ...677.1216H} 
for polytropes and by Lackey et al.~\cite{2012PhRvD..85d4061L} for the equations of state used by 
Kyutoku et al.~\cite{2012PhRvD..85d4061L}).
  
The predictions of these models typically vary by a few percents of the mass of the neutron star. From current data,
it is impossible to determine which one is most accurate. Differences in their predictions can however become larger 
outside of the fitting region, thus providing a useful estimate of our error. In the rest of this article, we will
consider numerical results from model~(\ref{eq:mod}) --- but as more simulations become available, and in particular
simulations with the same compactness $C_{\rm NS}$ but different equations of state, models~(\ref{eq:mod2}-\ref{eq:mod3})
might very well prove more accurate. 

\section{Numerical Results}
\label{sec:NR}

To fit the parameters $\alpha$ and $\beta$ of our model, we consider recent results from numerical relativity in the
range $q=3-7$, $\chi_{\rm BH}=0-0.9$ and $C_{\rm NS}=0.13-0.18$. We neglect simulations at lower
mass ratios, which are astrophysically less likely and cannot be modeled accurately by the simple formula assumed here. 
Larger spins and more compact stars would be interesting to consider: according to Hebeler et al.~\cite{2010PhRvL.105p1102H}, 
neutron stars of mass  $M_{\rm NS}\sim 1.4M_\odot$ could be in the range $C_{\rm NS}=0.15-0.22$, while
for the same neutron star mass, Steiner et al.~\cite{2010ApJ...722...33S}  find that the most likely compactness 
is $C_{\rm NS}=0.17-0.19$.  More massive stars should have an even higher compactness. 
As for the black hole spin, it is currently unconstrained --- and as
we will see, quasi-extremal black hole spins are a very interesting region of parameter space for BHNS mergers.

We also limit the model to spins aligned with the orbital angular momentum and to low-eccentricity orbits.
Misaligned spins have only been studied for one set of binary parameters~\cite{2011PhRvD..83b4005F}, so that we do not have
enough information about their influence on the disk mass to include them in the model.\footnote{
It is however worth noting that known precessing BHNS results, as well as soon-to-be published simulations
for higher mass ratios ($q=7$) and higher black hole spin ($\chi_{\rm BH}=0.9$) agree with the results
of our model if the radius of the innermost stable circular orbit is replaced by the radius of the 
innermost stable spherical orbit with the same inclination with respect to the black hole spin as the orbital plane of the binary, as proposed by Stone et al.~\cite{2012arXiv1209.4097S}.
}
High-eccentricity
mergers have been studied by East et al.~\cite{2011ApJ...737L...5S,2012PhRvD..85l4009E}, but again 
the data does not cover enough of the parameter space to be included in our fit. Additionally, eccentricity is only 
an issue for binaries formed in clusters: field binaries are expected to have negligible eccentricities at the time of merger.
Finally, we neglect the influence of magnetic fields, as both Etienne et al.~\cite{2012PhRvD..85f4029E} and 
Chawla et al.~\cite{2010PhRvL.105k1101C} find their effect on the remnant mass to be small 
(except for large interior magnetic fields $B\gtrsim 10^{17}G$).

A list of all simulations used to fit our model is given in Table~\ref{tab:sim}. These results were obtained by three
different groups: Kyoto~\cite{PhysRevD.84.064018} (SACRA code), UIUC~\cite{Etienne:2008re} and the SXS 
collaboration~\cite{2011PhRvD..83b4005F,2012PhRvD..85d4015F} (SpEC code). In those articles, the mass outside the black hole
$M^{\rm rem}_{\rm NR}$ is measured at different times, which would introduce a bias in our fit. We choose to use
the convention of Kyutoku et al.~\cite{PhysRevD.84.064018}, where $M^{\rm rem}_{\rm NR}$ is measured $10{\rm ms}$ after merger.
For this reason, the values listed in Table~\ref{tab:sim} differ from the masses given in the tables
of~\cite{Etienne:2008re,2011PhRvD..83b4005F,2012PhRvD..85d4015F}.

\begin{table}
\caption{Summary of the numerical results used. When more than one group simulated the same set of parameters, the average value
is used. $\chi_{\rm BH}=a_{\rm BH}/M_{\rm BH}$ is the dimensionless spin parameter of the black hole, 
$C_{\rm NS}=M_{\rm NS}/R_{\rm NS}$ is the compactness of the star, $M^{\rm rem}_{\rm NR}$ is the remaining mass $10{\rm ms}$ after
merger (as measured in the numerical simulations), and $M^b_{\rm NS}$ is the baryon mass of the star.}
\label{tab:sim}
\begin{tabular}{|c||c|c|c|c|c|c|}
\hline
ID & $\frac{M_{\rm BH}}{M_{\rm NS}}$ & $\chi_{\rm BH}$ & $C_{\rm NS}$ 
& $\frac{M^{\rm rem}_{\rm NR}}{M^b_{\rm NS}}$ & Code & Ref. \\ 
\hline
1 & 7 & 0.90 & 0.144 & 0.24 & SpEC & \cite{2012PhRvD..85d4015F} \\
2 & 7 & 0.70 & 0.144 & 0.05 & SpEC & \cite{2012PhRvD..85d4015F} \\
3 & 5 & 0.50 & 0.144 & 0.05 & SpEC & \cite{2012PhRvD..85d4015F} \\
4 & 3 & 0.90 & 0.144 & 0.35 & SpEC & \cite{2011PhRvD..83b4005F} \\
5 & 3 & 0.50 & 0.145 & 0.15 & SpEC/SACRA & \cite{2011PhRvD..83b4005F,PhysRevD.84.064018} \\
6 & 3 & 0.00 & 0.144 & 0.04 & UIUC/SpEC  & \cite{Etienne:2008re,2011PhRvD..83b4005F}\\
7 & 3 & 0.75 & 0.145 & 0.21 & UIUC/SACRA & \cite{Etienne:2008re,PhysRevD.84.064018} \\
8 & 5 & 0.75 & 0.131 & 0.25 & SACRA& \cite{PhysRevD.84.064018} \\
9 & 5 & 0.75 & 0.162 & 0.11 & SACRA& \cite{PhysRevD.84.064018} \\
10& 5 & 0.75 & 0.172 & 0.06 & SACRA& \cite{PhysRevD.84.064018} \\
11& 5 & 0.75 & 0.182 & 0.02 & SACRA& \cite{PhysRevD.84.064018} \\
12& 4 & 0.75 & 0.131 & 0.25 & SACRA& \cite{PhysRevD.84.064018} \\
13& 4 & 0.75 & 0.162 & 0.15 & SACRA& \cite{PhysRevD.84.064018} \\
14& 4 & 0.75 & 0.172 & 0.12 & SACRA& \cite{PhysRevD.84.064018} \\
15& 4 & 0.75 & 0.182 & 0.07 & SACRA& \cite{PhysRevD.84.064018} \\
16& 4 & 0.50 & 0.131 & 0.19 & SACRA& \cite{PhysRevD.84.064018} \\
17& 4 & 0.50 & 0.162 & 0.06 & SACRA& \cite{PhysRevD.84.064018} \\
18& 4 & 0.50 & 0.172 & 0.02 & SACRA& \cite{PhysRevD.84.064018} \\
19& 3 & 0.75 & 0.131 & 0.24 & SACRA& \cite{PhysRevD.84.064018} \\
20& 3 & 0.75 & 0.162 & 0.16 & SACRA& \cite{PhysRevD.84.064018} \\
21& 3 & 0.75 & 0.172 & 0.15 & SACRA& \cite{PhysRevD.84.064018} \\
22& 3 & 0.75 & 0.182 & 0.10 & SACRA& \cite{PhysRevD.84.064018} \\
23& 3 & 0.50 & 0.131 & 0.19 & SACRA& \cite{PhysRevD.84.064018} \\
24& 3 & 0.50 & 0.162 & 0.11 & SACRA& \cite{PhysRevD.84.064018} \\
25& 3 & 0.50 & 0.172 & 0.07 & SACRA& \cite{PhysRevD.84.064018} \\
26& 3 & 0.50 & 0.182 & 0.03 & SACRA& \cite{PhysRevD.84.064018} \\
\hline
27& 7 & 0.50 & 0.144 & 0.00 & SpEC & \cite{2012PhRvD..85d4015F} \\
28& 3 &-0.50 & 0.145 & 0.01 & UIUC & \cite{Etienne:2008re} \\
29& 5 & 0.00 & 0.145 & 0.01 & UIUC & \cite{Etienne:2008re} \\
30& 4 & 0.50 & 0.182 & 0.00 & SACRA& \cite{PhysRevD.84.064018} \\
31& 3 &-0.50 & 0.172 & 0.00 & SACRA& \cite{PhysRevD.84.064018} \\
\hline
\end{tabular}
\end{table}

Only some of the simulations listed in Table~\ref{tab:sim} were published with explicit error measurements. There is thus
some uncertainty on the accuracy of these results. From published convergence tests and our own experience with such
simulations, we assume that a rough estimate for the numerical errors $\Delta M^{\rm rem}_{NR}$ can be obtained by
combining a $10\%$ relative error and a $1\%$ absolute error in the mass measurement, i.e.
\beq
\label{eq:dMNR}\frac{\Delta M^{\rm rem}_{NR}}{M^b_{\rm NS}}=\sqrt{ \left(\frac{0.1 M^{\rm rem}_{\rm NR}}{M^b_{\rm NS}}\right)^2 + 0.01^2 }.
\eeq

A few of the parameter sets from Table~\ref{tab:sim}  have been studied by multiple groups (ID 5,6,7). 
It should be noted however that these simulations are actually different cases: the compactness of the
neutron star is similar for all groups ($C_{\rm NS}=0.144$ for SpEC, $C_{\rm NS}=0.145$ for UIUC 
and $C_{\rm NS}=0.146$ for SACRA), but the equations of state used are quite different (SpEC and 
UIUC use a $\Gamma=2$ polytrope, while the results from SACRA were obtained with a piecewise
polytrope with different internal structure). Even so, the results are compatible with the error estimates~(\ref{eq:dMNR})
[i.e. differences $\sim 0.01-0.03M_{\rm NS}$].
The values listed in Table~\ref{tab:sim} are averages of the numerical results of the
different groups.

\section{Parameter Estimates}
\label{sec:params}

\subsection{Best-Fit parameters}

We determine the parameters $\alpha$ and $\beta$ of our model (Eq.\ref{eq:mod}) through a least-square fit for the results
of simulations 1-26 in Table~\ref{tab:sim}. Simulations 27-31, which do not lead to the formation of a disk,
are not used directly --- but we check that the model is consistent with their results. We find
\beqn
\alpha &=& 0.288 \pm 0.011 \label{eq:Pa}\\
\beta  &=& 0.148 \pm 0.007\label{eq:Pb},
\eeqn
for model $M^{\rm rem}_{\rm model}$ in which tidal forces are estimated from Newtonian physics, and
\beqn
\tilde \alpha &=& 0.296 \pm 0.011\\
\tilde \beta  &=& 0.171 \pm 0.008
\eeqn
for the modified model $\tilde M^{\rm rem}_{\rm model}$ in which the tidal forces are derived from the Kerr metric.

Error estimates are easier if we rewrite the models using singular value decomposition (see e.g p65-75 and p793-796 
of Press et al.~\cite{numrec_cpp}, and references thererin), that is if we transform the basis functions of our model 
so that the parameters of the model have uncorrelated errors. For example, in the case of the 'Newtonian' model we have
\beqn
f_1 &=& 0.851 \left(3q\right)^{1/3}(1-2C_{\rm NS}) - 0.525 \frac{R_{\rm ISCO}}{R_{\rm NS}} \nonumber\\
f_2 &=& 0.525 \left(3q\right)^{1/3}(1-2C_{\rm NS}) + 0.851 \frac{R_{\rm ISCO}}{R_{\rm NS}} \nonumber
\eeqn
\beq
\frac{M^{\rm rem}_{\rm model}}{M^b_{\rm NS}} = Af_1+Bf_2.
\eeq
The best-fit parameters $A$ and $B$ are then
\beqn
A &=& 0.323 \pm 0.013 \\
B &=& 0.026 \pm 0.001,
\eeqn
where the errors on $A$ and $B$ are independent (while the errors on $\alpha$ and $\beta$ were strongly correlated).
 
\subsection{Goodness-of-fit}

The ability of these models to fit the numerical results within their errors $\Delta M^{{\rm rem}}_{NR}$ can be
estimated through the reduced $\chi^2$
\beq
\chi^2=\frac{1}{N^{df}}\Sigma_{i=1}^{26}\left(\frac{M^{{\rm rem},i}_{\rm model}-M^{{\rm rem},i}_{\rm NR}}
{\Delta M^{{\rm rem},i}_{NR}}\right)^2
\eeq
where $N^{df}=26-N_{\rm params}=24$ is the number of degrees of freedom, and the index $i$ refers to the ID of the numerical 
simulations (i.e. $M^{{\rm rem},1}$ is the remnant mass for simulation 1 of Table~\ref{tab:sim}, and $\Delta M^{{\rm rem},1}_{NR}$ the
corresponding error estimate computed from Eq.~[\ref{eq:dMNR}]). The 'Newtonian' model $M^{\rm rem}_{\rm model}$ and the 'Kerr' model $\tilde M^{\rm rem}_{\rm model}$ are 
equally good fit to the data, with $\chi^2=0.98$ and $\chi^2=0.96$ respectively.
By comparison, the best-fit results for model 
$M^0_{\rm model}$ (in which we do not correct $d_{\rm tidal}$ by the factor $[1-2C_{\rm NS}]$) has a much larger $\chi^2=4.04$.
Adding a constant term $\gamma$ to either $M^{\rm rem}_{\rm model}$ or $\tilde M^{\rm rem}_{\rm model}$ leads to $\chi^2=1.00$.

\begin{figure}
\caption{Predictions of the best-fit model (diamonds) for simulations 1-26.
The solid line represents the ideal $M^{\rm rem}_{\rm model}=M^{\rm rem}_{\rm NR}$ result, 
while the error bars correspond to the estimated numerical errors $\Delta M^{\rm rem}_{\rm NR}$.}
\label{fig:errM}
\includegraphics[width=8.3cm]{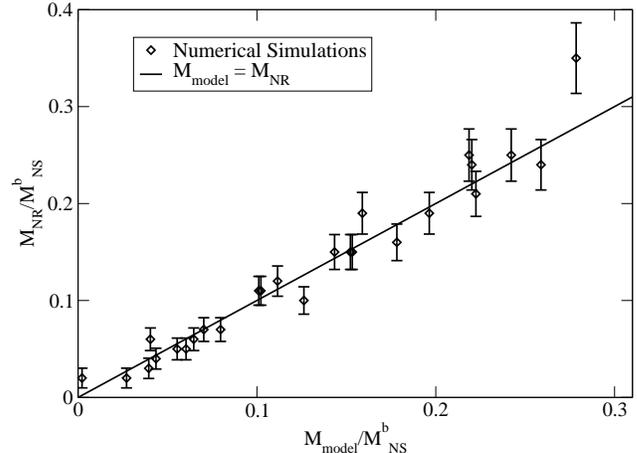}
\end{figure}

A comparison between the simple model $M^{\rm rem}_{\rm model}$ and the numerical results is shown in Fig.~\ref{fig:errM}, 
in which we plot $M^{\rm rem}_{\rm NR}$ as a function of $M^{\rm rem}_{\rm model}$ for simulations 1-26. 
We can see that the difference between the modelled and measured masses is generally smaller than the errors expected
from Eq.~\ref{eq:dMNR}. The main exception is the large remnant mass observed in case 4. We suspect that our model, which assumes
that the remnant mass scales linearly with $R_{\rm ISCO}$ and $d_{\rm tidal}$, breaks down for remnant masses
greater than about $20-25\%$ of the neutron star mass. A non-linear relation between these distances and the remnant mass might
perform better in that regime, but more numerical simulations are required to test that hypothesis. 

The more complex model $\tilde M^{\rm rem}_{\rm model}$ offers very similar results: for cases 1-26, the worst disagreement between 
the models is $0.008M^b_{\rm NS}$ (for case 3) while their rms difference is $0.004M^b_{\rm NS}$. 
Models~(\ref{eq:mod2}-\ref{eq:mod3}) show larger variations, of order of a few percents of the neutron star mass.

\subsection{Error Estimates}
\label{sec:error}

Estimating the error in the mass predictions of our model from the statistical errors in the parameters $\alpha$ and $\beta$
is likely to be misleading. Differences between the numerical results $M^{\rm rem}_{\rm NR}$ and the predictions of the model
$M^{\rm rem}_{\rm model}$ come from multiple sources: the numerical error $\Delta M^{\rm rem}_{\rm NR}$ of course, but also
a physical spread of the exact mass remnants around the predictions of the model. 
A part of that spread at least should be due to differences in the outcome of BHNS mergers for binaries with the same parameters 
($M_{\rm BH}$,$M_{\rm NS}$,$\chi_{\rm BH}$,$C_{\rm NS}$), but different equations of state (i.e. neutron stars with the same
radius but a different internal structure). This effect can also be seen in the differences between the predictions of 
models~(\ref{eq:mod},\ref{eq:mod2},\ref{eq:mod3}). But more generally, it is unlikely that the simple equations used here can perfectly 
represent the complex dynamics of a BHNS merger. 

From the fact that we measured $\chi^2 \sim 1$, we know that the errors $M^{\rm rem}_{\rm model}-M^{\rm rem}_{\rm NR}$ are 
compatible with a gaussian distribution of variance $\Delta M^{\rm rem}_{\rm NR}$. This is already indicative of the likely 
existence of a non-zero physical spread around the results of the model. 
The estimated numerical errors $\Delta M^{\rm rem}_{\rm NR}$ are
indeed more of an upper-bound on the errors in the simulations than the width of an expected gaussian distribution. In
the absence of a difference between the real physical outcome of a merger and the output of the model, 
we would thus expect $\chi^2$ to be
lower than 1. How much of the measured errors $M^{\rm rem}_{\rm model}-M^{\rm rem}_{\rm NR}$ comes from numerical errors and how
much from actual differences between the model and the physical reality is hard to determine, especially considering
that the numerical errors are not well known. A more cautious approach to estimate the uncertainty in the model is thus to 
consider $\Delta M^{\rm rem}_{\rm NR}$ as a conservative upper bound on the variance of a gaussian error in the model itself.

\begin{figure}
\caption{$M^{\rm rem}_{\rm model}=0.1M^b_{\rm NS}$ contours for, from top to bottom, 
neutron star compactness $C_{\rm NS}=0.22,0.18,0.155,0.135$  (i.e. $R_{\rm NS}\approx 9.5,11.5,13.5,15.5{\rm km}$ 
for $M_{\rm NS}=1.4M_\odot$). For each compactness, we have $M^{\rm rem}_{\rm model}>0.1M^b_{\rm NS}$ above the plotted contour. 
The shaded regions encompass the portions of phase space for which $M^{\rm rem}_{\rm model}=0.1M^b_{\rm NS} \pm \Delta M^{\rm rem}_{NR}$. 
SGRBs are extremely unlikely to occur below the green region ($C_{\rm NS}=0.155$). 
Note that the scale is chosen in order to zoom on the high-spin region (the y-axis scales as $\log(1-\chi_{\rm BH})-\log(\chi_{\rm BH})$).}
\label{fig:MC10}
\includegraphics[width=8.3cm]{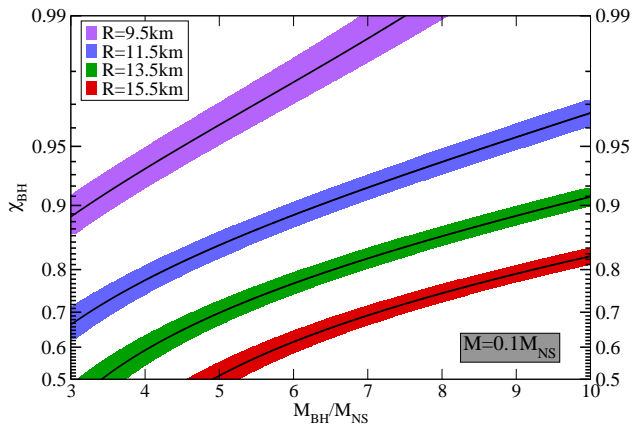}
\end{figure}

Fig.~\ref{fig:MC10} shows contours of $M^{\rm rem}_{\rm model}=0.1M^b_{\rm NS}$ for various neutron star compactness. 
The general features of this plot are not surprising: the formation 
of massive disks is known to be favored by low mass ratios, high black hole spins and large neutron stars. 
But our model allows for the determination of the region of parameter space in which a certain amount of matter will remain 
available at late times with fairly high accuracy: at least within the spread $\Delta M^{\rm rem}_{NR}\approx 0.02M_\odot$ 
or, if we consider a measurement of $M^{\rm rem}$ as a way to determine the radius of a neutron star, 
within $\Delta R_{\rm NS}\lesssim 0.5{\rm km}$.

Another important issue is the validity of the model outside the parameter range currently covered by numerical relativity.
It is indeed possible that larger errors will be found for more compact neutron stars ($C_{\rm NS} > 0.18$) or larger mass ratios
($M_{\rm BH} > 7 M_{\rm NS}$). However, given that our model fits the numerical data over a fairly wide range of parameters, 
and is derived from the physics of tidal disruption, it is likely to give decent results over most of the 
astrophysically relevant parameter space --- with the notable exception of configurations leading to very large 
remnant masses $M^{\rm rem} \gtrsim 0.20-0.25 M_{\rm NS}$ (i.e. for nearly-extremal black hole spins and low mass ratios), and
probably of the asymptotic regime $\chi_{\rm BH}\rightarrow 1$ where scalings valid in the range $\chi=0-0.9$ might
break down. The differences between models~(\ref{eq:mod},\ref{eq:mod2},\ref{eq:mod3}) outside of the fitting
region can also provide a rough estimate of these errors.

\section{Discussion}
\label{sec:discussion}

\subsection{Parameter Space Study}

The models described in the previous sections can be used to easily approximate the region of parameter space in which
disruption occurs, or in which a certain amount of mass will remain available at late times. Such predictions are particularly 
important when trying to determine which BHNS mergers are likely to lead to short gamma-ray bursts (SGRBs): only 
BHNS mergers ending with the formation of a massive accretion disk could power SGRBs. Disruption of the neutron star
is also a prerequisite for the ejection of unbound material, and thus for any electromagnetic signal due to the
radioactive decay of a neutron-rich ejecta.
If the neutron star does not disrupt,
the only observational signatures of BHNS mergers are their gravitational wave emissions, as well as potential 
electromagnetic or neutrino precursors (see e.g. Tsang et al.~\cite{2012PhRvL.108a1102T}).

The minimum remnant mass required to get SGRBs is
currently unknown, and is likely to vary across the parameter space: the fraction of the remnant mass which, at any given
time, is in a long-lived accretion disk around the black hole (as opposed to the tidal tail or unbound ejecta)
is by no means a constant, nor are the physical characteristics of that disk. 
We know, for example, that at high mass ratios a larger fraction of the mass is initially 
in an extended tidal tail than for lower mass black holes~\cite{2012PhRvD..85d4015F}. 
Furthermore, other characteristics of the disk (temperature, thickness, baryon loading along the rotation axis of the
black hole, magnetic fields) are important for the generation of a gamma-ray burst. And what the ideal conditions
are depends on the physical process powering the burst (see Sec.~\ref{sec:sgrb} for more details).
Nevertheless, $M^{\rm rem}_{\rm model}$ is already a useful prediction,
providing a good estimate of the amount of material available for post-merger evolution.
Additionally, any configuration for which $M^{\rm rem}_{\rm model}=0$ can be immediately rejected as a potential SGRB
progenitor. 

\begin{figure}
\caption{Contours $M^{\rm rem}_{\rm model}=(0,0.05,0.1,0.15,0.2)M^b_{\rm NS}$ for a star of compactness 
$C_{\rm NS}=0.155$ ($R_{\rm NS}\approx 13.5{\rm km}$ for $M_{\rm NS}=1.4M_\odot$).
The shaded regions correspond to portions
of parameter space in which no matter remains around the black hole (bottom/red), more than $0.2M^b_{\rm NS}$ remains and 
massive disks should be the norm (top/green), and an intermediate region in which lower mass disks will form (center/blue).
Note that the scale is chosen in order to zoom on the high-spin region.}
\label{fig:C155}
\includegraphics[width=8.3cm]{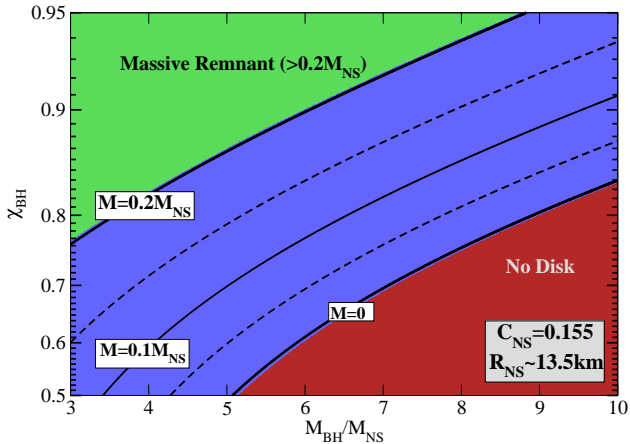}
\end{figure}

\begin{figure}
\caption{Same as Fig.~\ref{fig:C155}, but for $C_{\rm NS}=0.18$ ($R_{\rm NS}\approx 11.5{\rm km}$).}
\label{fig:C18}
\includegraphics[width=8.3cm]{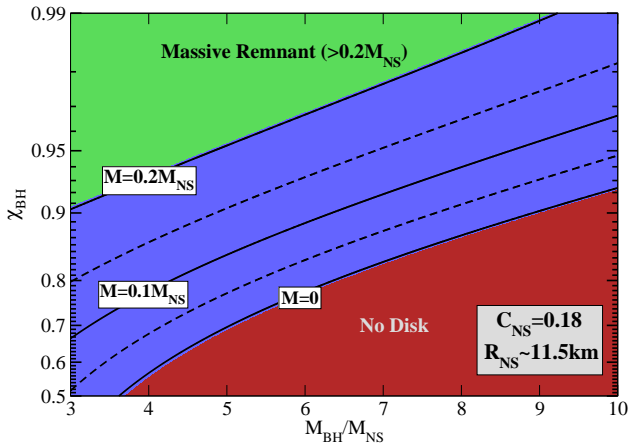}
\end{figure}

\begin{figure}
\caption{Same as Fig.~\ref{fig:C155}, but for $C_{\rm NS}=0.22$ ($R_{\rm NS}\approx 9.5{\rm km}$).}
\label{fig:C22}
\includegraphics[width=8.3cm]{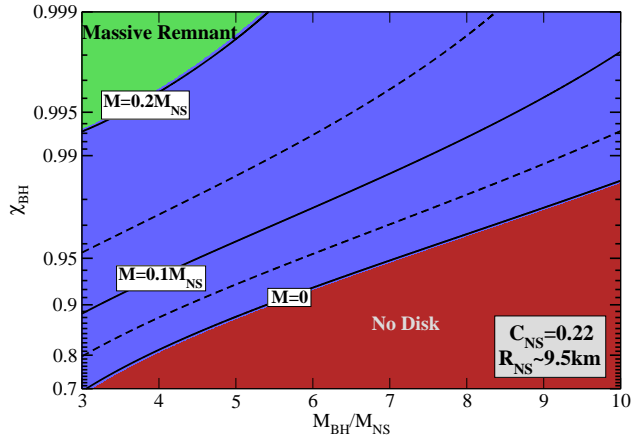}
\end{figure}

Predictions for the mass of neutron star material remaining outside the black hole at late times are detailed in 
Figs.~\ref{fig:C155}-\ref{fig:C22},
in which we plot contours of the remnant mass as a function of the mass ratio and black hole spin. Each figure correspond
to a different neutron star compactness, covering the range of radii expected from the theoretical results of Hebeler et 
al.~\cite{2010PhRvL.105p1102H}. 
Experimental measurements of neutron star radii are still fairly difficult, but studies of bursting X-ray binaries by 
Ozel et al.~\cite{2009ApJ...693.1775O,2010ApJ...719.1807G,2012ApJ...748....5O} tend to favor the lower range of potential radii 
($R_{\rm NS}\approx 9-12$ km). Steiner et al.~\cite{2010ApJ...722...33S}, after reassessing
the errors in the measurement of neutron star radii from X-ray bursts, derived a parametrized equation of state 
which takes into account both the astrophysical measurements and results from nuclear theory. They predict that 
$R_{\rm NS}\approx 11-12$ km for $M_{\rm NS}=1.4M_\odot$. We can thus consider Fig.~\ref{fig:C155} and Fig.~\ref{fig:C22}
as bounding the range of potential neutron star radii, while Fig.~\ref{fig:C18} is around the most likely neutron star
size (for $1.4M_\odot$ stars --- heavier stars are expected to be more compact).

The strong dependence of the remnant mass in the radius of the star is particularly
noteworthy. In the most likely astrophysical range of mass ratios ($q\sim 5-10$), remnant
masses $M^{\rm rem}=0.1M_{\rm NS}$ can be achieved for moderate black hole spins 
$\chi_{\rm BH}\approx 0.7-0.9$ if we consider neutron stars with $C_{\rm NS}=0.155$ ($R_{\rm NS}\approx 13.5{\rm km}$), as
in Fig.~\ref{fig:C155}. But at the other end of the range of potential neutron star radii, 
for $C_{\rm NS}=0.22$ ($R_{\rm NS}\approx 9.5{\rm km}$), the much more restrictive condition $\chi_{\rm BH} \approx 0.9-0.999$ 
applies(Fig.~\ref{fig:C22}). 
For a neutron star in the range of compactness favored by Steiner et al.~\cite{2010ApJ...722...33S} 
($C_{\rm NS}=0.18$, or $R_{\rm NS}\approx 11.5{\rm km}$), keeping $10\%$ of the neutron star material outside 
the black hole requires spins $\chi_{\rm BH}\approx 0.8-0.97$, an already fairly restrictive condition (Fig.~\ref{fig:C18}). 

This naturally implies that the rate of SGRBs produced as a result of BHNS mergers is very sensitive to the equation 
of state of nuclear matter, and in particular to the size of neutron stars. Determining that rate is unfortunately 
impossible without knowledge of the number of BHNS mergers, and of the distributions of black hole spins and mass ratios. 
Additionally, a large enough $M^{\rm rem}$ is only a necessary condition for a given BHNS binary to power a SGRB. Knowledge
of the exact properties of the accretion disk (and of the exact physical process leading to short gamma-ray bursts) would be
required to accurately determine which BHNS systems are SGRB progenitors.
Nonetheless, the importance of the equation of state can be fairly easily understood by simply 
computing the area of the region above a certain contour of $M^{\rm rem}_{\rm model}$ for various values of $C_{\rm NS}$. 
Let us define $\chi^c(M,C_{\rm NS},q)$ as the critical spin above which $M^{\rm rem}_{\rm model}>M$ and
\beq
\phi(M,C_{\rm NS})=\frac{\int_5^{10}[1-\chi^c(M,C_{\rm NS},q)]{\rm dq}}{5}.
\eeq
Then, $\phi(M,C_{\rm NS})$ represents the fraction of binaries with mass remnants greater than $M$ assuming that the 
distributions of mass ratios and spins are uniform within the $q=5-10$ and $\chi_{\rm BH}=0-1$ range respectively. 
As we decrease the size of the neutron star from $C_{\rm NS}=0.155$ to $C_{\rm NS}=0.22$, Table~\ref{tab:rate} shows
that we go from about $20\%$ of the parameter space in which significant disks are possible to about $1\%$! This does not 
necessarily mean that SGRBs are impossible for $C_{\rm NS}\sim 0.22$ --- but certainly indicate that they would occur in a 
non-negligible  fraction of BHNS mergers only if quasi-extremal spins are the norm.

\begin{table}
\caption{Fraction $\phi(M,C_{\rm NS})$ of the parameter space within $q=5-10$, $\chi_{\rm BH}=0-1$ for which 
$M^{\rm rem}_{\rm model}>M$ for various neutron star compactness $C_{\rm NS}$ and critical masses $M$.}
\label{tab:rate}
\begin{tabular}{|c||c|c|c|}
\hline
$C_{\rm NS}$ & $\phi(0,C_{\rm NS})$ & $\phi(0.1M^b_{\rm NS},C_{\rm NS})$ & $\phi(0.2M^b_{\rm NS},C_{\rm NS})$  \\ 
\hline
0.135 & 0.46 & 0.30 & 0.16 \\
0.155 & 0.29 & 0.17 & 0.07 \\
0.180 & 0.16 & 0.08 & 0.02 \\
0.220 & 0.05 & 0.01 & 0.00 \\
\hline
\end{tabular}
\end{table}

\begin{figure}
\caption{Contours of $M^{\rm rem}_{\rm model}$ for binaries with mass ratio $M_{\rm BH}=7M_{\rm NS}$ ($M_{\rm BH}\approx 10M_\odot$).  
Shown are contours for $M^{\rm rem}_{\rm model}=(0,0.05,0.1,0.15,0.2)M^b_{\rm NS}$. The shaded regions are as in Fig.~\ref{fig:C155}
and the neutron star radius (top scale) is computed assuming a star of ADM mass $M_{\rm NS}=1.4M_\odot$.}
\label{fig:Q7}
\includegraphics[width=8.3cm]{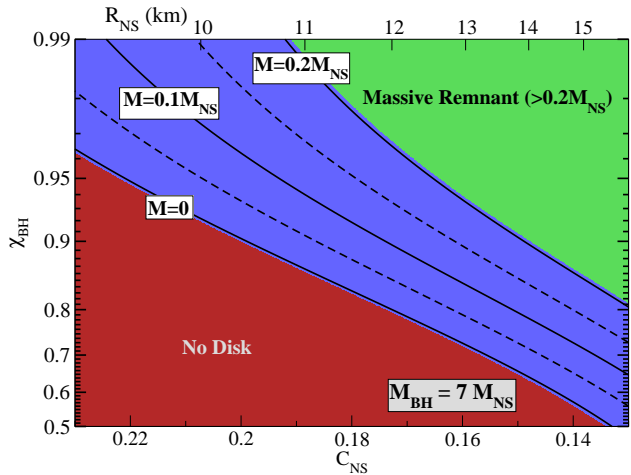}
\end{figure}

Current population synthesis models estimate the peak of the distribution of black hole masses in BHNS systems to
be around $M_{\rm BH}\sim 10M_\odot$, or $M_{\rm BH}\sim 7 M_{\rm NS}$~\cite{2008ApJ...682..474B,2010ApJ...715L.138B}. 
Fig.~\ref{fig:Q7} offers clearer information
on the behavior of BHNS systems in that regime. We see that no disk can form for $\chi_{\rm BH}<0.9$ unless 
$R_{\rm NS}>10.5{\rm km}$. That condition becomes $R_{\rm NS}>12 {\rm km}$ if we require at least $0.1M_{\rm NS}^b$ 
outside the black hole at late times. Results for BHNS binaries with higher black hole spins 
($\chi_{\rm BH}\rightarrow 1$) should of course be considered with caution: 
indeed, no mergers of BHNS binaries 
have been published for $\chi_{\rm BH}>0.9$ or $C_{\rm NS}>0.18$, and such simulations would be required to
rigorously test the accuracy of these predictions in extreme regions of the parameter space. Nonetheless, our model
indicates that quasi-extremal spins are at least a necessary condition for the formation of massive disks for 
$M_{\rm BH}\approx 10M_\odot$ and  $R_{\rm NS}\leq 12{\rm km}$. 

\begin{figure}
\caption{Contours $M^{\rm rem}_{\rm model}=0.1M^b_{\rm NS}$ for varying black hole spins
$\chi_{\rm BH}=(0.5,0.7,0.9,0.99)$. The grey region contains spins above the maximum value reached by numerical
simulations.}
\label{fig:CA}
\includegraphics[width=8.3cm]{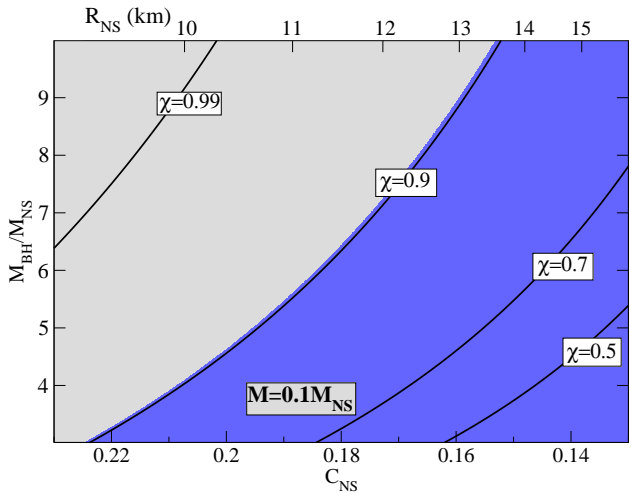}
\end{figure}

The minimum spin requirement for massive disk formation across
the parameter space of BHNS binaries is shown in Fig.~\ref{fig:CA}, in which the black hole spin needed to 
keep $10\%$ of the neutron star mass outside the black hole at late times is plotted. Figs.~\ref{fig:Q7} and~\ref{fig:CA}
both indicate the existence of an extended region of parameter space ($C_{\rm NS}\sim 0.18-0.22$, $\chi_{\rm BH}\sim 0.9-1$) 
which is likely to be astrophysically relevant but remains numerically unexplored,
and in which the outcome of BHNS mergers varies significantly.

\subsection{Comparison with previous models}
\label{sec:prevmod}

The tidal disruption of a BHNS binary is a complex problem, to which various approximations have been proposed.
In the limit of very large black hole masses, Fishbone~\cite{1973ApJ...185...43F} derived the separation at which
equilibrium tides would cause the disruption of an incompressible,  corotating neutron star, a work that was generalized
to compressible flows and irrotational binaries by Wiggins \& Lai~\cite{2000ApJ...532..530W}. Such models have a few
significant limitations, which were discussed in more details by Miller~\cite{2005ApJ...626L..41M}.
The innermost stable circular orbit of the black hole is only an approximation to the minimum separation at which
stable circular orbits exist for finite mass objects.
Analytical approximations to the location of the last stable circular orbit can be obtained from the Post-Newtonian
expansion (see e.g. Damour et al.~\cite{2000PhRvD..62h4011D}). These show that for equal mass objects
the last stable circular orbit can be well outside of the ISCO obtained in the point particle limit. An alternative
to the analytical method, which avoids the complications resulting from the use of the Post-Newtonian expansion
close to merger, is to consider sequences of quasi-equilibrium configurations computed numerically. This is the
approach taken by Taniguchi et al.~\cite{2008PhRvD..77d4003T} to determine whether a neutron star in a BHNS
binary would disrupt before reaching the last stable circular orbit (in the case of non spinning black holes). The
numerical results also indicate that the innermost stable orbit is outside of the ISCO of the isolated 
black hole, although not by as much as the Post-Newtonian results would indicate.

Miller~\cite{2005ApJ...626L..41M} also points out that the condition used 
in~\cite{2000PhRvD..62h4011D,2008PhRvD..77d4003T} is only valid in the limit of infinitely slow inspiral. If the system
looses angular momentum through the emission of gravitational waves, the plunge will actually begin outside of
the last stable circular orbit, thus limiting further the ability of BHNS binaries to disrupt and form accretion disks. 
Additionaly, models based on equilibrium tides neglect the fact that, close to disruption, the rapid inspiral can cause the
neutron star to be well out of equilibrium.

An alternative method is to simply fit a semi-analytical models to the result of numerical relativity, effectively attempting
to include the complex physics that is not taken into account by the model into the free parameters of the fit. This is the
approach taken by Pannarale et al.~\cite{2011ApJ...727...95P}, in their model describing the neutron star as a
tri-axial ellipsoid distorted by the tidal field of the black hole. That model was however fitted to general relativistic
simulations at low mass ratio which have since been shown to have underestimated in many cases the mass of
the remnant. At high mass ratio, no general relativistic simulations were available at the time, and the model was
thus fitted to simulations using an approximate treatment of gravity, known to overestimate the ability of BHNS binaries
to form disks. Our model takes a similar approach, fitting a rather simple physical model to more recent numerical
data covering a wider range of binary parameters.

Compared to the predictions of Pannarale et al.~\cite{2011ApJ...727...95P}, the results presented here indicate that it is a lot 
more difficult to create massive accretion disks at high mass ratios than what that previous model indicated, while
at low mass ratios $q\sim 3$ our model predicts significantly higher final masses. 
As opposed to~\cite{2011ApJ...727...95P}, our model is unlikely to capture the behavior of BHNS mergers
with $q\sim1$, when finite-size effects begin to make it more difficult to form massive disks. These differences
are expected considering what we now know of the limitations of the numerical data used to fit their model.

We can also revisit the condition derived by Taniguchi et al.~\cite{2008PhRvD..77d4003T} for the parameters allowing
disk formation in the case of non-spinning black holes, and by Wiggins \& Lai~\cite{2000ApJ...532..530W} for extreme mass ratios. 
Requiring $M_{\rm model}^{\rm rem}>0$ is equivalent to imposing an upper bound on the neutron star compactness,
\beq
\label{eq:Clim}
C_{\rm NS}\lesssim \left(2+2.14q^{2/3}\frac{R_{\rm ISCO}}{6M_{\rm BH}}\right)^{-1}.
\eeq
We find that our results are less favorable to tidal disruption and disk formation than in~\cite{2008PhRvD..77d4003T}, 
as could be expected from the arguments of Miller~\cite{2005ApJ...626L..41M} discussed at the beginning of this
section.
The predictions of Eq.~(\ref{eq:Clim}) are on the other hand in agreement with the numerical simulations performed by 
Kyutoku et al.~\cite{2010PhRvD..82d4049K}, even though the results for low mass ratio, non-spinning BHNS mergers 
published in~\cite{2010PhRvD..82d4049K} were not taken into account when fitting our model. In the high mass ratio
limit, they also agree fairly well with the results of Wiggins \& Lai~\cite{2000ApJ...532..530W} 
(within $\sim 15\%$ for $q\sim 1000$).
This is however more of a test of the ability of our model to extrapolate to extreme mass ratios, well outside of the
fitting region, than of the accuracy of the results of Wiggins \& Lai, which are more reliable in that regime.

\section{Conclusions}

We constructed a simple model predicting the amount of matter remaining outside the black hole about $10{\rm ms}$ after
a BHNS merger, based on comparisons between the binary separation at which the neutron star is expected to be 
disrupted by tidal forces from the black hole and the radius of the innermost stable circular orbit around the hole. 
We show that the model can reproduce the results of recent general relativistic simulations of non-precessing, 
low-eccentricity BHNS mergers within a few percents of the total mass of the neutron star. The simplest best-fit model is
\beq
\frac{M^{\rm rem}}{M^b_{\rm NS}} \approx 0.288 \left(3\frac{M_{\rm BH}}{M_{\rm NS}}\right)^{1/3}\left(1-2\frac{M_{\rm NS}}{R_{\rm NS}}\right) 
- 0.148 \frac{R_{\rm ISCO}}{R_{\rm NS}} \nonumber
\eeq
(in units in which $G=c=1$).

These mass predictions should be valid at the very least within
the range of parameters currently covered by numerical simulations ($M_{\rm BH}=3-7M_{\rm NS}$, $R_{\rm NS}=11-16$ km, 
$a_{\rm BH}/M_{\rm BH}=0-0.9$), and are likely to remain fairly accurate within most of the astrophysically relevant parameter
space. Alternative models using different approximations for the binary separation at which tidal disruption
occurs are presented in Sec.~\ref{sec:model}.

Using this model, it becomes easy to estimate the region of parameter space in which large amounts of matter remain outside
the black hole for long periods of time. This is of particular importance when studying whether BHNS mergers can
result in short gamma-ray bursts. Our results show the strong dependence of the remnant mass in the radius of the
neutron star: whether the equation of state of neutron stars is at the soft or stiff end of its potential range could easily 
translate into an order of magnitude difference in the rate of gamma-ray bursts originating
from BHNS mergers. It is also quite clear that high black hole spins are 
likely to be a prerequisite for the formation of 
massive disks. Neutron stars in the middle of the theoretically allowed range of radii ($R_{\rm NS}\sim 11.5{\rm km}$) 
require spins $a_{\rm BH}/M_{\rm BH} \gtrsim 0.8$ for about 10\% of the neutron star material to remain outside the hole, 
while quasi-extremal spins are necessary for the most compact stars.

The validity of our model is currently limited to black hole spins aligned with the orbital angular momentum and
remnants below $\sim 20-25\%$ of the neutron star mass, due to the lack of numerical data available for precessing binaries
and high mass remnants. Extending the model to cover these interesting parts of the parameter space would certainly be
useful, but would require a large number of computationally intensive simulations to be performed (particularly to cover
misaligned black hole spins). A few additional simulations using high mass ratios or small neutron star radii together with
relatively large spins ($\chi_{\rm BH}\gtrsim 0.9$) would also be extremely helpful, allowing better estimates of the errors 
in the model for binary parameters which are astrophysically relevant but have never been considered by numerical relativists. 

The extreme simplicity of these models should make them useful tools to obtain cheap but reliable estimates of the results 
of BHNS mergers across most of the astrophysically relevant parameter space, as well as to help determining which numerical 
simulations to perform in order to study given physical effects. This simplicity is, however, also a reason for caution: to 
accurately predict which BHNS systems would lead to the production of short gamma-ray bursts, modeling more physical
properties will certainly be required: temperature, division of the mass between disk and tidal tail, neutrino emission, and magnetic
field configuration are all important characteristics of the final remnant, as are the final properties of the black hole
recently modeled by Pannarale~\cite{2012arXiv1208.5869P}. More detailed models, however, might require a larger number
of numerical simulations as the number of fitted parameters and the complexity of the problem increases.
Finally, an improved understanding of the physical process leading to a burst will also be necessary before we can explicitly determine which BHNS binaries produce short gamma-ray bursts.

\acknowledgments

The author thanks Harald Pfeiffer, Saul Teukolsky, Dong Lai, Christian Ott and Matthew Duez for useful 
discussions and suggestions concerning this project, and all members of the SXS collaboration for regular 
discussions and input. 
The author also wish to thank the participants of the 'Rattle and Shine' conference at the Kavli Institute
for Theoretical Physics for stimulating discussions on this topic. This research was supported in part by the National
Science Foundation under Grant No.NSF PHY11-25915.

\bibliography{References/References}

\end{document}